\documentclass{article}
\usepackage{graphicx}
\usepackage[utf8]{inputenc}

\def\arcsinh{\mathop{\mbox{arcsinh}}}
\def\diag{\mathop{\mbox{diag}}}

\begin{document}
\title{Stability of white holes revisited}
\author{Igor Nikitin\\
Fraunhofer Institute for Algorithms and Scientific Computing\\
Schloss Birlinghoven, 53757 Sankt Augustin, Germany\\
\\
igor.nikitin@scai.fraunhofer.de
}
\date{}
\maketitle

\begin{abstract}
It is shown that the models of white hole interacting with external matter can be made stable by introduction of a negative central mass. Similar results are obtained for the models of white hole, interacting with null shells, with radial flows of matter and with photon gas. In realistic models, a naked timelike singularity corresponding to the negative mass is hidden under a coat of positive mass, providing an extremely strong redshift for photons born in the Planck neighborhood of the singularity and observed at infinity, thereby realizing the principle of cosmic censorship in a relaxed form.
\end{abstract}

\begin{figure}
\begin{center}
\includegraphics[width=\textwidth]{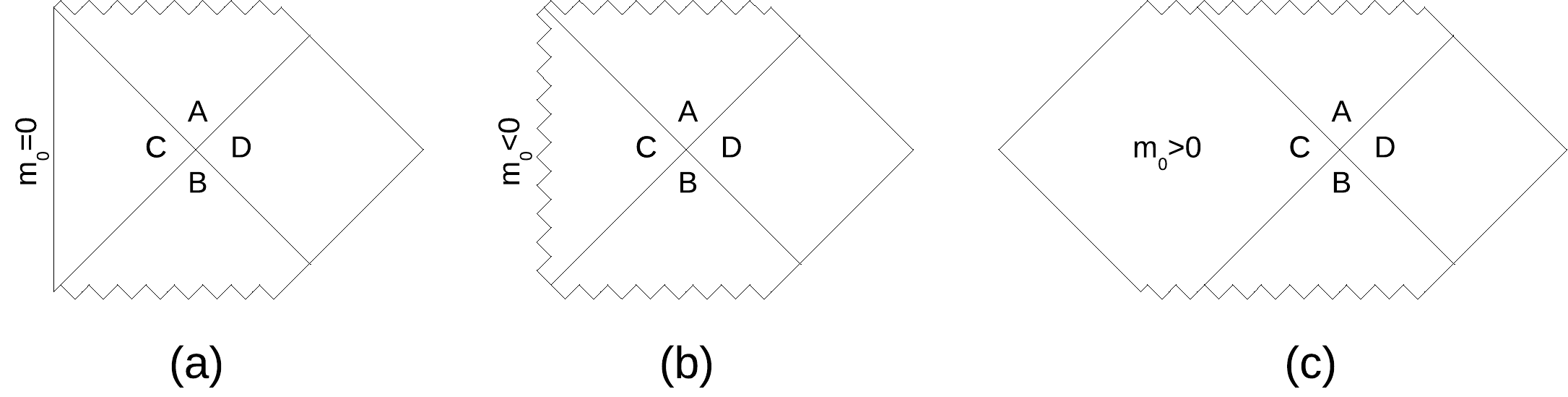}
\end{center}
\caption{Null shell model of white hole. On Penrose diagrams the crossing of two null shells separates the space-time into four regions marked ABCD. Three special cases are considered for the region C: (a) zero central mass, (b) negative central mass, (c) positive central mass.}\label{f1}
\end{figure}

\section{Introduction}

It is generally believed that the white holes, T-symmetrical analogs of the black holes, become unstable when they start interacting with external matter, see the papers by Eardley~\cite{Eardley}, Ori and Poisson~\cite{OriPoisson}, Barceló et al.~\cite{151100633}. The simplest way to this conclusion is to consider the model of a white hole with outgoing and ingoing flows of matter, represented as null shells. Problems with stability begin at the moment when the null shell, representing the flow of matter ejected by the white hole, meets the null shell, representing the flow of matter falling to the white hole from the outside. When these shells cross in the vicinity of the white hole's particle horizon, their energy is redistributed. As a result, the ingoing shell receives a greater part of the energy, while only a negligible part of the initial energy of the outgoing shell comes through to the outside. Finally, the ingoing shell goes under its own event horizon, thereby making the white hole black, emitting virtually nothing to the outer space.

In this paper we are going to reconsider some aspects of this model. First of all, modeling of external matter, usually representing relic radiation or scattered light of stars, in the form of collapsing shell, can be questioned. The boundary condition in the form of collapsing shell clearly violates T-symmetry. If we consider T-conjugated boundary condition in the form of an expanding shell, then stability problems will appear not for white, but for black holes. Instead, if we consider external matter as photon gas or null fluid, it will be T-symmetric. In a photon gas for any group of photons from which a collapsing shell can be formed, there is a group of photons that make up the expanding shell.

Collapsing null shell also violates spatial isotropy and homogeneity. The photons in the shell move toward the center, even at a great distance, when the action of the central mass does not manifest itself. That is, this motion is not associated with the gravitational attraction of the central mass, but is due to a special choice of initial conditions. Also, in collapsing null shell there is no tangential pressure, there is only radial one. In a photon gas, the distribution of photons at any point is isotropic and there are both radial and tangential pressure components.

The collapsing shell preserves its total energy, and while its area shrinks with decreasing radius, the energy density increases inversely with the square of the radius. Further, if the photons are directed exactly to the center, the shell, as a result of its own evolution, even in the absence of the central mass, will go under the event horizon and form a black hole. On the other hand, the rarefied photon gas is approximately homogeneous, with the exception of small thermal fluctuations, and spontaneous production of black holes does not occur there.

Another possible extension of the model is the introduction of a negative mass. Such a modification, of course, contradicts the energy conditions, usually imposed on the solutions of general relativity. On the other hand, the energy conditions have been repeatedly criticized, see the paper by Barceló and Visser~\cite{ec-crit} and references therein. Negative masses are required in many interesting astrophysical models, such as wormholes, various versions presented by Visser in his book~\cite{Visser-wrmh-book}, and warp drives, by Alcubierre~\cite{Alcubierre-warp-drive}. The fundamental work on the interaction of null shells by Dray and 't~Hooft~\cite{Dray-tHooft} has also considered the cases with negative mass.

In our previous work~\cite{static-rdm} T-symmetric stationary scenario with radial matter flows has been studied. These flows can be represented as a continuous sequence of ingoing and outgoing shells. At the center of the model there is a  singularity of ultraviolet type, possessing negative mass. In the other our work~\cite{wrmh-rdm}, the negative mass in the model was distributed in space, leading to the opening of a wormhole in the center. T-symmetry and stationarity, characteristic to these models, give them the properties both of white and black holes, which absorb matter from the outer universe, throw it out and are in a dynamic equilibrium with it.

In this paper, we will continue to study T-symmetric stationary scenarios in the model of white and black holes, interacting with external matter flows. In Sec.\ref{sec2}, we will show that the introduction of a negative mass allows to stabilize the model of a white hole with collapsing null shell. As a result, the white hole can throw out an arbitrary fraction of its mass, depending on the choice of model parameters. In particular, it can throw out the amount of energy equal to the energy of collapsing shell, thereby realizing T-symmetric scenario. In Sec.\ref{sec3}, radial matter flows will be considered in T-symmetric stationary scenario previously investigated in \cite{static-rdm}. We will calculate Misner-Sharp mass in this model and show that in the center of the system there is a core of negative mass. In Sec.\ref{sec4}, a photon gas described by TOV equation will be considered and T-symmetric stationary solutions will be investigated, possessing negative central mass. 

\section{Interaction of white hole with null shell}\label{sec2}

This scenario is shown in Fig.\ref{f1}. Two null shells cross each other and divide the spacetime into four areas, marked ABCD. In each of these four regions a vacuum solution with own mass is established:
\begin{equation}
m_A=M-E,\  m_B=M-dm,\  m_C=m_0,\  m_D=M.
\end{equation}
Here we follow the notation of Ori and Poisson \cite{OriPoisson}: $M$ -- the mass of a white hole with a collapsing shell, $dm$ -- the mass of the collapsing shell, $E$ -- the energy of the outgoing shell measured at infinity. The difference is that for the region C we consider an arbitrary mass $m_0$, remaining in the center of the white hole after the radiation of the outgoing shell, while Ori and Poisson considered the case of zero mass and flat spacetime in region C, Fig.\ref{f1}a. We also allow for this mass both positive and negative values. For negative mass, Fig.\ref{f1}b, spacetime in region C has the form of Schwarzschild solution with a naked timelike singularity. For positive mass, Fig.\ref{f1}c, the global structure of spacetime has a form of Kruskal-Szekeres maximal extension of Schwarzschild solution, the eternal black hole. Note that in \cite{OriPoisson} other global structure was used that links vacuum solutions with a cosmological model. As we will see, this does not affect the formulae expressing energetic characteristics of the white hole.

Next, we use Dray-'t Hooft-Redmount (DTR) relation \cite{Dray-tHooft}
\begin{equation}
f_Af_B=f_Cf_D,\ f_i=1-2m_i/R,\ i=A,B,C,D.
\end{equation}
Here $R$ is the radius value at collision point and the system of units $G=c=1$ is used. Solving this equation for $E$ and introducing new variables
\begin{equation}
\xi=R/(2M)-1,\ \alpha=dm/M,\ \beta=m_0/M,\ \eta=E/M,
\end{equation}
we have the expression for an efficiency of white hole explosion:
\begin{equation}
\eta=(1 - \alpha - \beta)\,\xi/(\alpha + \xi).
\end{equation}
At $\beta=0$ it is identical with (2.3) in \cite{OriPoisson}. Note that at $\beta=0$ and $0< \xi \ll \alpha \ll 1$ we have $\eta\sim\xi/\alpha \ll 1$, i.e., the closer is the collision of null shells to the particle horizon, the smaller fraction of the mass of the white hole is radiated by it to the infinity.

Further, let's consider the evolution of the null shell: $\xi=\xi_0\exp(-t/(2M))$, where $t$ is the time of distant observer, see, e.g., (25.30) in Blau, Lecture Notes on General Relativity~\cite{Blau}. Summing the time necessary for ingoing shell to reach the collision point and the time necessary to outgoing shell to escape from the collision point to the starting distance, the time is doubled: $\tau=2t$ and we have
\begin{equation}
\xi=\xi_0\exp(-\tau/(4M)).\label{xitau}
\end{equation}
Here $\xi_0$ represents the starting distance relative to the Schwarzschild's radius. It should be sufficiently small to keep the approximation under (25.30) \cite{Blau} valid, however, it has only a gauge meaning and a little influence to the final result. The total time necessary for null shell to come from a large distance to the collision point is summed up from (i) the time from a large distance to the $\xi_0$-vicinity of particle horizon, this time is approximately proportional to the starting distance and is controlled by it, and (ii) the time of waiting at the particle horizon, which the distant observer will measure for the propagation of the null shell from $\xi_0$ to the collision point $\xi$, located even closer to the particle horizon. For practically relevant settings of the problem, this latter time prevails, moreover, at $\xi\to0$ the term $\log\xi$ dominates over $\log\xi_0$ in the expression for $\tau=-4M(\log\xi-\log\xi_0)$.

These formulae are sufficient to reproduce the estimations for efficiency of white hole explosion obtained by Eardley \cite{Eardley} and Ori and Poisson \cite{OriPoisson}. Indeed, fixing $\beta=0$ and considering the limit $0<\xi,\alpha \ll1$, we have $\eta=(1+\alpha/\xi)^{-1}$, identical with (2.4) \cite{OriPoisson}. Further, using (\ref{xitau}), we see that $\xi=\alpha$, i.e., $\eta=0.5$ is reached at $\alpha=\xi_0\exp(-\tau_d/(4M))$, i.e., $\tau_d=-4M\log(\alpha/\xi_0)$. Thus, we have $\eta=(1+\alpha/\xi_0\exp(\tau/(4M)))^{-1}=(1+\exp((\tau-\tau_d)/(4M)))^{-1}$, identical with (3.15) \cite{OriPoisson}. Further, considering small $\alpha$ and moderate $\xi_0$, we obtain an estimation $\tau_d\sim4M\log(M/dm)$, identical with \cite{Eardley}.

The main result of \cite{Eardley} and \cite{OriPoisson} is that considering the time values of the order of the age of the universe and the masses of the white hole much smaller than the mass of the observable universe, one will have large values $\tau/(4M)\gg1$. It leads to exponentially small values of $\xi$ and $\eta$. Numerically, they are extremely small indeed, of the order $\xi,\eta\sim\exp(-10^{5})$ in \cite{OriPoisson}. This estimation corresponds to the distances much less than Planck length or any other lower limit on distances used in physics. The reason is the exponential evolution $\xi\sim\exp(-\tau/(4M))$, producing so small numbers for large $\tau$-values. Evidently, this estimation is an extrapolation in the range of small distances, while the main result is qualitative, the radiated energy is proportional to $\xi$ and is extremely small. If $\xi$ will be cut off at Planck length, the efficiency will still be enormously small.

If negative mass values are allowed, new opportunities are opening up. For $\beta<0$ and $0< \xi \ll \alpha \ll 1$ the value $\eta\sim(1-\beta)\,\xi/\alpha$ can be $\sim1$, if $\beta\sim -\alpha/\xi$. Considering exact formulae, we see that $\eta=0.5$, i.e., half-efficient white hole explosion with $E=M/2$, can be reached at
\begin{equation}
\beta=-(\alpha - \xi + 2 \alpha \xi)/(2 \xi),
\end{equation}
while $\eta=\alpha$, T-symmetric case of $E=dm$, can be reached at
\begin{equation}
\beta = -(\alpha^2 - \xi + 2 \alpha \xi)/\xi.
\end{equation}

The reason for the appearance of these solutions is that the white hole can emit the energy greater than its original mass, leaving the negative mass behind. The greater part of this energy will return back and revert this mass to positive. At least, such behavior is encoded in the scenario Fig.\ref{f1}b, where the timelike singularity is linked with two spacelike ones. Further, only a small part of the initial outgoing energy comes through. The point is that the initial outgoing energy can be made arbitrarily large, thus for the final outgoing energy in absolute units the large values can be also obtained.

Formally, let's define a new efficiency $\eta_2=E/(M-dm-m_0)$, the final energy of outgoing shell relative to the initial energy of outgoing shell, rather than to the total mass of the system. After simplifications we have
\begin{equation}
\eta_2=\xi/(\alpha+\xi).
\end{equation}
It is interesting, that this expression coincides with (2.4) \cite{OriPoisson}, however, the efficiency there is $E/M$ and this expression appears only in the limit $0<\xi,\alpha \ll1$, while here it is the new type of efficiency $\eta_2$ and the relation is exact. Remarkably, it does not depend on $\beta$, i.e., $m_0$. Therefore, at $0<\xi\ll\alpha\ll1$ the value $\eta_2\sim\xi/\alpha$ can be extremely small, but for obtaining $E$ in absolute units it is multiplied to $(M-dm-m_0)$, which can be made arbitrarily large by choosing the negative $m_0$ values.

\begin{table}
\begin{center}
\caption{Various scenarios in null shell model of the white hole}\label{tab1}

~

\begin{tabular}{|c|c|c|c|c|}\hline
$M/M_\odot$& $\xi=R/(2M)-1$& $\tau=4M\log(\xi_0/\xi)$& $m_0/M_\odot$& $\eta=E/M$\\ \hline
$4.06\cdot10^6$& $\sim\exp(-10^{16})$& $13.8\cdot10^9$~years& $0$& $~99\xi$\\ 
$4.06\cdot10^6$& $10^{-5}$& $737$~sec& $0$& $10^{-3}$\\ 
$4.06\cdot10^6$& $\sim\exp(-10^{16})$& $13.8\cdot10^9$~years& $-2\cdot10^4/\xi$& $0.5$\\ 
$4.06\cdot10^6$& $\sim\exp(-10^{16})$& $13.8\cdot10^9$~years& $-4\cdot10^2/\xi$& $0.01$\\ 
$4.06\cdot10^6$& $10^{-5}$& $737$~sec& $-2.03\cdot10^9$& $0.5$\\ 
$4.06\cdot10^6$& $10^{-5}$& $737$~sec& $-3.67\cdot10^7$& $0.01$\\ 
$10$& $10^{-5}$& $1.81$~msec& $-5\cdot10^3$& $0.5$\\ 
$10$& $10^{-5}$& $1.81$~msec& $-90.2$& $0.01$\\ \hline
\end{tabular}

~

(common parameters: $\alpha=dm/M=0.01$, $\xi_0=0.1$) 
\end{center}
\end{table}

In Table~\ref{tab1}, we calculated several interesting scenarios for the null shell model of the white hole. First, we consider the mass of a white hole of the order of the central galactic black hole. In the first line of the table we take the time of the order of the age of the universe and set the central mass $m_0$ to zero. For definiteness, the parameters $\alpha=0.01$, $\xi_0=0.1$ are fixed, but the final result depends little on them. For $\xi$ and $\eta$ the extremely small values $\sim\exp(-10^{16})$ are obtained. In work \cite{OriPoisson}, in a similar setting with a different choice of model parameters, the extremely small values $\sim\exp(-10^{5})$ were also obtained.

In the second line, in order to show that this result is entirely determined by the choice of the parameter $\tau$, a moderately small $\xi=10^{-5}$ value is chosen, corresponding to $\tau=737$~sec. As a result, the moderately small efficiency value $\eta=10^{-3}$ is obtained.

In the next four lines, we show that the choice of a negative mass allows one to obtain predetermined values of efficiency, in particular, the case of half-efficiency and the T-symmetric case with $\eta=\alpha$. 

For time $\tau$ of the order of the age of the universe, the extremely large absolute values of the mass are obtained. These values significantly exceed the mass of the observable universe and arise from the extremely small values of $\xi$-distance, to which they are inversely proportional. As we discussed above, the application of the model in the range of such small distances is the extrapolation, and the result should be understood qualitatively. Cutoff of $\xi$ at the Planck length also leads to a very large negative mass in the center. For moderately small values of $\xi$, moderately large values of negative mass are obtained.

The last two lines represent the case of the white hole mass in the order of stellar black hole mass, with the similar results.

In this paper, we consider the case when, according to the clock of an external observer, the shells collide in a finite time, which is possible only for $R>2M$. The case considered by Eardley \cite {Eardley} and Barceló et al. \cite {151100633} is also possible, corresponding to slightly different Penrose diagrams. This is the case when the outgoing shell does not come out at all, from the viewpoint of the external observer, but collides with the incoming shell under the event horizon. For the external observer, what happens under the horizon is inaccessible, and this case is indistinguishable from that, as if the outgoing shell did not exist at all, that is, the solution would look like an eternal black hole, onto which an additional shell was thrown. In the present work, our goal is to find the regions of model parameters in which the explosion of the white hole occurs, principally observable from the outside.

\section{Interaction of white hole with radial flows of matter}\label{sec3}

\begin{figure}
\centering
\includegraphics[width=0.8\textwidth]{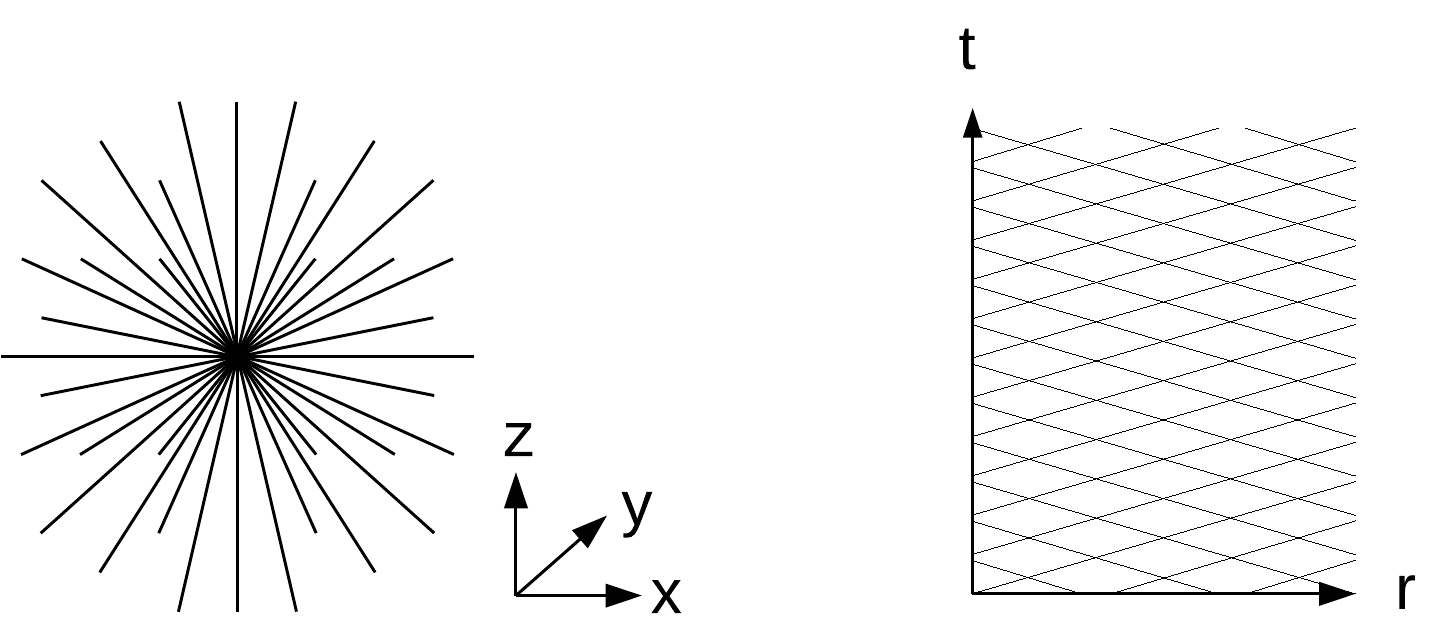}
\caption{Static spherically symmetric problem with radially directed flows of matter, in $xyz$-coordinates on the left, in $tr$-coordinates on the right. Image from~\cite{static-rdm}.}
\label{f2a}
\end{figure}

In this section we will consider a static model with T-symmetric ingoing and outgoing radially directed matter flows, shown on Fig.\ref{f2a}. This matter distribution was used in \cite{static-rdm} as a model for dark matter halo in spiral galaxies. Here we will consider a special case of light-like (null directed) matter flows. Also, it will not be so important to us whether this matter is dark or not. It interests us as a model of a compact massive object, permanently emitting and absorbing the matter flows. We will set a boundary condition in the form of converging and diverging shells at infinity and will integrate the solution inwards, to have a look what happens there.

We use spherical coordinates $(t,r,\theta,\phi)$ and a standard metric
\begin{equation}
ds^2=-A(r)dt^2+B(r)dr^2+r^2(d\theta^2+\sin^2\theta\; d\phi^2)\label{stdmetr}
\end{equation}
and select energy-momentum tensor
\begin{equation}
T^{\mu\nu}=\rho(r)(u_+^\mu(r)u_+^\nu(r)+u_-^\mu(r)u_-^\nu(r)),\label{Tmunu}
\end{equation}
where $\rho(r)$ is the density and $u_\pm(r)=(\pm u^t(r),u^r(r),0,0)$ are the velocity fields of outgoing and ingoing radial matter flows. 

The general solution of geodesic equations has a form: 
\begin{eqnarray}
&&4\pi\rho=c_1/\left(r^2u^r\sqrt{AB}\right),\ u^t=c_2/A, \ u^r=\sqrt{c_2^2+c_3A}/\sqrt{AB},\label{eq_geode}
\end{eqnarray}
with positive constants $c_{1,2}$ and the third constant defining a norm $c_3=u_\mu u^\mu$. Three variants $c_3=0,\pm1$ for this norm have been considered in \cite{static-rdm}, while here we will consider only one case: $c_3=0$, null radial dark matter (NRDM). Further difference with \cite{static-rdm} is the appearance of $4\pi$ factor in $\rho$-equation, coming due to a different system of units used in the present paper, $G=c=1$.

\begin{figure}
\begin{center}
\includegraphics[width=0.45\textwidth]{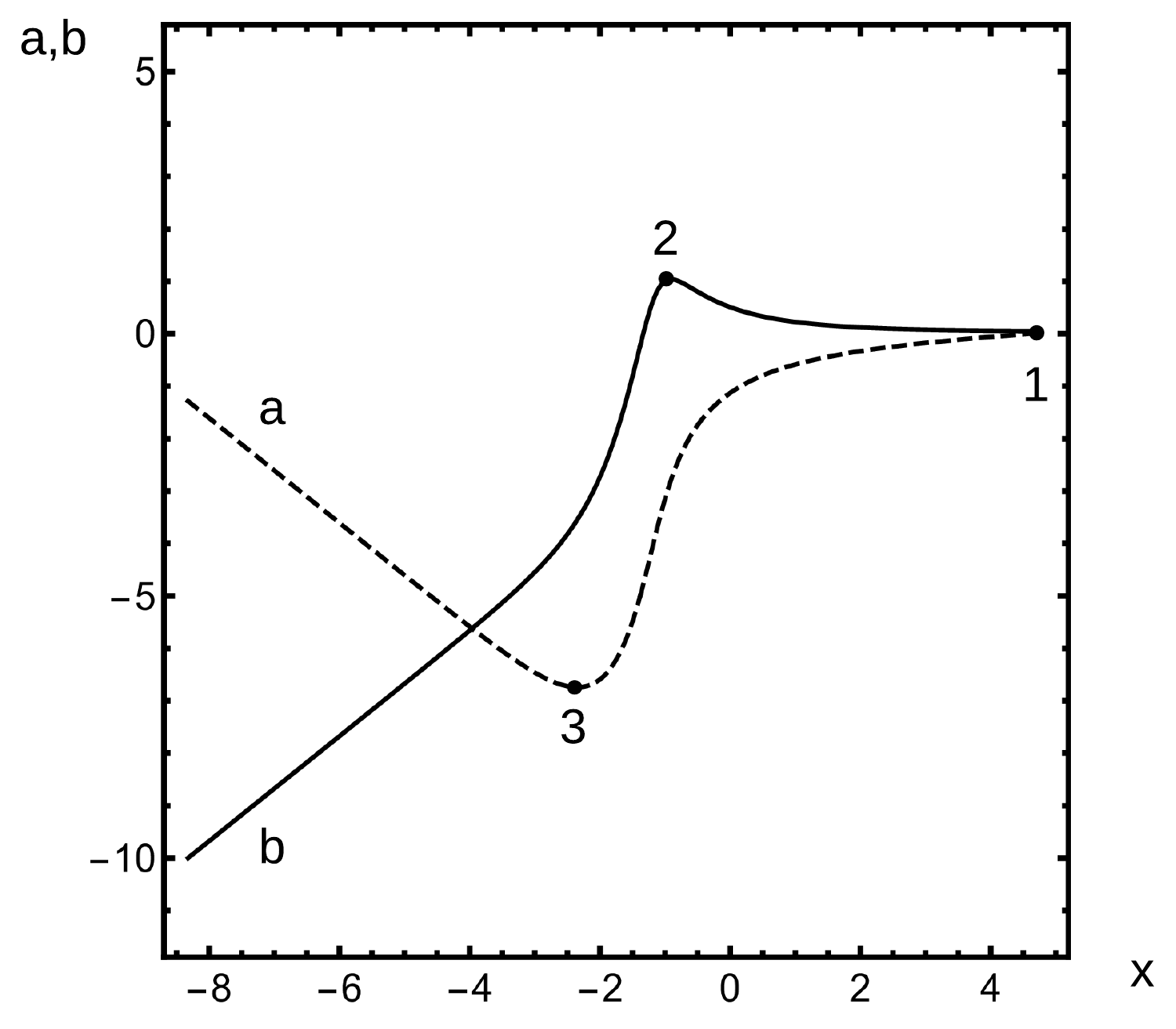}
~~\includegraphics[width=0.45\textwidth]{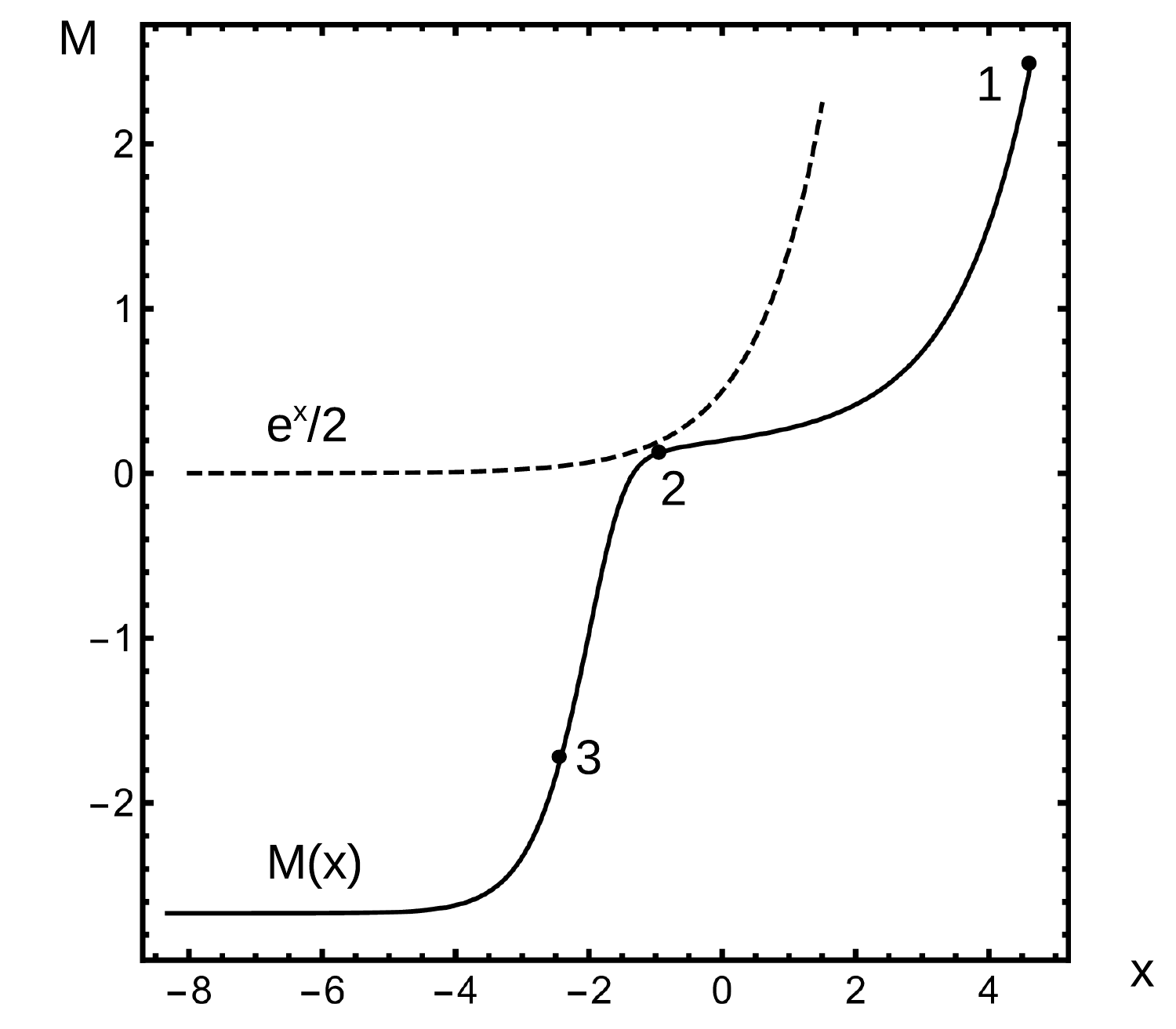}
\end{center}
\caption{On the left: NRDM solution with $\epsilon=0.04$, the evolution of $a=\log A$, $b=\log B$ as functions of $x=\log r$. Starting from point~1, the solution reaches maximal $b$ in the point~2. Then the functions $a$ and $b$ fall down rapidly until the solution reaches minimal $a$ in the point~3. Then the functions $a,b$ go apart symmetrically. On the right: Misner-Sharp mass for the solution. Starting from $M\sim\epsilon r/2$ near the point~1, the solution comes close to the horizon $M\sim r/2$ in the point~2 and is bounced into the negative region. There it comes through the inflection point~3 and tends to a negative constant at $r\to0$.}\label{f2}
\end{figure}

The field equations to solve are
\begin{eqnarray}
&&da/dx=-1+e^b+\epsilon\ e^{b-a},\label{eq_dadx}\\
&&db/dx=1-e^b+\epsilon\ e^{b-a},\label{eq_dbdx}
\end{eqnarray}
in logarithmic variables
\begin{eqnarray}
&&a=\log A,\ b=\log B,\ x=\log r,\ \epsilon=4c_1c_2,\label{eq_c45}
\end{eqnarray}
with the starting point
\begin{eqnarray}
&&x_1=\log r_1,\ a_1=0,\ b_1=\epsilon + r_s/r_1,\label{xab1}
\end{eqnarray}
where $r_s$ is a gravitational radius of the central massive object, $r_1$ is the starting radius of integration. We solve this system numerically and show the result on Fig.\ref{f2} left.

Setting $\epsilon=0.04$ and $r_s=1$, we start integration from a point~1, located well above $r_s$. The solution reaches a maximum of $b$ in a point~2, after that the functions fall down almost in parallel. In a point~3 a minimum of $a$-function is reached and the functions go apart with opposite slopes.

On the right part of Fig.\ref{f2} we have presented Misner-Sharp mass \cite{Blau} for this solution, which in our coordinates reads
\begin{eqnarray}
&&M=r/2\ (1-B^{-1})=e^x/2\ (1-e^{-b}).
\end{eqnarray}
At large $r$ the function $M(r)$ is almost linear with a small slope, corresponding to a dark matter halo in the model of spiral galaxies. A difference $M(r_1)-M(r_2)$ represents a mass of the halo. Further, at smaller $r$, the solution shows Schwarzschild's behavior and goes towards $M>r/2$ zone, where black or white holes are formed. The solution comes very close to this zone, however, the formation of horizon does not happen. The solution is bounced into a region of negative masses, where it tends to a negative constant at $r\to0$. The mass of dark matter in this model is always positive, while Misner-Sharp mass, a total mass enclosed by a sphere of radius $r$, becomes negative due to the contribution of the central singularity. The difference $M(r_2)-M(0)$ represents a positive mass of a coat surrounding a singularity of negative mass $M(0)$.

In more details, the radial density of Misner-Sharp mass is given by $M'(r)=4\pi r^2\rho_{\mbox{\footnotesize eff}}$, where $\rho_{\mbox{\footnotesize eff}}$ is an effective mass density of dark matter flows, i.e., a component of energy-momentum tensor $T_\nu^\mu=\mbox{diag}(-\rho_{\mbox{\footnotesize eff}},p_{\mbox{\footnotesize eff}},0,0)$. In \cite{static-rdm} the general expression for this density is given, which for NRDM case and our system of units becomes 
\begin{eqnarray}
&&\rho_{\mbox{\footnotesize eff}}=p_{\mbox{\footnotesize eff}}=\epsilon/(8\pi r^2A).\label{rhopeff}
\end{eqnarray}
This density is positive, leading to positive contribution of dark matter to Misner-Sharp mass. 

The asymptotic formulae for different zones in the solution are available in \cite{static-rdm}. The asymptotic solution at large $x$ is 
\begin{eqnarray}
&&a=Const+2\epsilon x -r_s e^{-x},\ b=\epsilon+r_s e^{-x},\label{eq_z1}
\end{eqnarray}
leading to Misner-Sharp mass $M=(\epsilon r +r_s)/2$. On the other hand, using $\varphi=a/2$ as Newtonian potential, one obtains the expression for the orbital velocity $v$ and the orbital acceleration 
\begin{eqnarray}
&&v^2/r=\varphi'_r=\epsilon/r +r_s/(2r^2).\label{eq_ar}
\end{eqnarray}
At large $r$ the dark matter term $\epsilon/r$ dominates over Newtonian term $r_s/(2r^2)$, producing asymptotically flat rotation curves $v^2=\epsilon$. One can also notice that this orbital acceleration corresponds to a mass function $M_{\mbox{\footnotesize grav}}=\epsilon r +r_s/2$ with the dark matter contribution twice larger than its Misner-Sharp mass. The reason is that the pressure also contributes to gravity and for the considered NRDM case the contributions of the mass density $\rho_{\mbox{\footnotesize eff}}$ and the pressure $p_{\mbox{\footnotesize eff}}$ are equal.

Further, the solution comes close to the horizon $M=r/2$. The relative distance to the horizon line $1-M/(r/2)=e^{-b}$ is minimal in the maximum of $b$, i.e., in the point 2. After passing the point 2, the functions $a,b$ start to fall rapidly, reaching there large negative values. This phenomenon leads to extremely large redshift for this region as observed from the large distances and has been termed in \cite{static-rdm} as {\it red supershift}. Through (\ref{rhopeff}) extremely large negative $a$ provides extremely large positive values of density and pressure, leading to the formation of dark matter coat with the extremely fast mass accumulation. A similar phenomenon, {\it mass inflation} for counterstreaming matter flows, has been observed by Hamilton and Pollack in the paper \cite{HamiltonPollack}, studying the structure of charged black holes. 

In point 3 the $a$-function reaches a minimum, corresponding to a maximum of radial density $M'(r)$, or, equivalently, inflection point of $M(r)$. Due to the logarithmic transform, it is formally different from the inflection point of $M(x)$, although visually the graph $M(x)$ is also consistent with the inflection point located in the range of point 3.

After passing the point 3, the radial density gradually falls down. The coat continues till the origin, where the solution enters the regime typical for naked Schwarzschild's singularity:
\begin{eqnarray}
&&a=-x+c_{12},\ b=x+c_{13},\label{eq_z3}
\end{eqnarray}
with large negative constants $c_{12,13}$. It corresponds to a large negative constant mass $M=-e^{-c_{13}}/2$, located in the origin.

Note that for the case of massive dark matter (MRDM) considered in \cite {static-rdm} there is {\it a vacuole} around the central singularity, due to the presence of turning points on massive geodesics, which arise due to the gravitational repulsion of the central negative mass. For NRDM case, there is no vacuole, the coat closely joins the singularity, but its radial density begins to decrease after passing the point 3, also due to the repulsion of the central singularity. This repulsive effect resembles a quantum mechanical bounce discussed in the work by Barceló et al.~\cite {151100633} and the references therein. Although, in our model the bounce is classical, appearing due to the gravitational repulsion of the central singularity.

With $\epsilon\to0$, the solution outside the gravitational radius arbitrarily closely approaches the Schwarzschild's solution, however, under the gravitational radius, in this limit, the effects of supershift and mass inflation become stronger and stronger. Thus, the solution from the outside fits arbitrarily close to a black or white hole, but is different from the inside. Such a structure of massive compact objects was also found in other models, see the review by Visser et al. \cite{09020346}.

Now we will calculate the key points of NRDM model for the parameters typical for the dark matter halo in our Milky Way galaxy. The calculation is similar to \cite {static-rdm}, it uses a different choice of the constants associated with the type of matter. For compatibility, we write the complete set of constants defined in \cite{static-rdm}, for our special case:
\begin{eqnarray}
&&c_1=\epsilon/4,\ c_2=1,\ c_3=c_5=0,\ c_4=c_6=c_7=\epsilon. 
\end{eqnarray}
Most of the model constants are bound to $\epsilon$, related to the orbital velocity $v$ on asymptotically flat rotation curves. The measurements for the Milky Way galaxy in the paper by Sofue and Rubin \cite{SofueRubin} give $v\sim200$~km/s, corresponding to $\epsilon=(v/c)^2\sim4\cdot10^{-7}$. The estimations for the mass of the central black hole in our galaxy have been also made by Ghez et al. \cite{08082870}, corresponding to $r_s\sim1.2\cdot10^{10}$m. Now we have all parameters necessary to start the integration. The results are presented in Table~\ref{tab2}.

\begin{table}
\begin{center}
\caption{NRDM model parameters and ranges for the Milky Way galaxy}\label{tab2}

~

\def\arraystretch{1.1}
\begin{tabular}{|c|c|}
\hline
model parameters&$\epsilon=4\cdot10^{-7}$, $r_s=1.2\cdot10^{10}$m\\ \hline
a border of the galaxy,&$r_1=3.1\cdot10^{21}$m,\\ 
starting point of&  $a_1=0$, $b_1=4.00004\cdot10^{-7}$,\\ 
the integration& $\log_{10}(M_1/M_\odot)=11.6231$\\ \hline
&$r_{2}=1.07883\cdot10^{10}$m, $a_{2}=-14.7318$,\\ 
supershift begins&$b_{2}=13.3455$, $\log_{10}(M_2/M_\odot)=6.56265$,\\
& $r_2-2GM_2/c^2=17261.9$m\\ \hline
&$r_{3}=6.79592\cdot10^6$m, \\
supershift ends&$a_{3}=b_3-14.7238$, $b_{3}=-1.24995\cdot10^6$,\\ 
& $\log_{10}(-M_3/M_\odot)=5.4285\cdot10^5$\\ \hline
minimal radius& $r_4=1.62\cdot10^{-35}$m, \\ 
(Planck length),&$a_4=a_3+95.3439$, $b_4=b_3-96.3359$,\\
end of the integration& $M_4=1.64\ M_3$\\ \hline
\end{tabular}

\end{center}
\end{table}

At the outer limit of the galaxy, at $r_1=100$~kpc from the center, the integration is started, and the clock for the global time are set, $a_1=0$. Misner-Sharp mass for the system, including the central massive object and the dark matter halo, at this point is $M_1=4.2\cdot10^{11} M_\odot$. 

At $r_2=1.08\cdot10^{7}$~km, approximately 16 solar radii, $b$-function reaches a maximum, then a supershift regime typical for RDM model begins. The value $r_2$ is located a bit below the nominal value $r_s$ for gravitational radius. 
Misner-Sharp mass at this point, including only the central object, is $M_2=3.653\cdot10^6 M_\odot$. The object is very close to a formation of horizon, the difference of the actual radius $r_2$ and the gravitational radius for the mass $M_2$ is only $17$~km, a small value in comparison with $r_2$ itself.

Further, at $r_{3}=6.8\cdot10^3$~km, approximately Earth radius, the supershift regime ends. The values $a_{3}\sim b_{3}=-1.25\cdot10^6$ are reached there. Misner-Sharp mass at this point is deeply negative, {\it its logarithm} is $\log_{10}(-M_3/M_\odot)\sim5\cdot10^5$. This value is similar to the extremely large negative masses for the central object obtained in the previous section.

In further decrease of the radius, $a$ increases and $b$ decreases according to the naked singularity asymptotics (\ref{eq_z3}). However on the lower limit at Planck length, where we stop the integration, the absolute variation of $a$ and $b$ is of the order $10^2$, much less than the values themselves, of the order $10^6$. In particular, the redshift defined by $a$-value is still extremely large at this point. Misner-Sharp mass is increased by $64\%$ relative to the point 3, rather steadily than the exponential inflation on the previous stage.

As a variant of the calculation, we also considered the case when, at the outer radius of the galaxy, the boundary condition is set not to the typical density of dark matter, but to the density of the relic radiation. The density value on the outer radius becomes 7 orders less. Further, the geometry of the solution assumes that the density will then increase inversely to the square of the radius, by 22 orders of magnitude from $ r_1 $ to $ r_2 $, which, of course, is not correct for relic radiation. This phenomenon does not occur because of the influence of the black hole, but only because of the choice of initial conditions with the concentration of light rays at the center of the system. More correct behavior will be obtained in the next section when considering a photon gas. However, we performed the calculation for this scenario as well. Reducing $ \epsilon $ generally leads to sharper dependencies, making the calculation a challenge for the integrator. The graphs slide further into the region of negative $ x $, while the point 3 is shifted behind the Planck length. Therefore, we concentrate here on the scenario with the typical density of dark matter, as a more representative case.

There is another version of the calculation with the modification of the model at the minimum radius. In \cite {wrmh-rdm} we have shown that the central singularity in the RDM solution can be relatively easily replaced with a wormhole. Both solutions contain negative mass, but the throat of the wormhole is characterized by $ B \to \infty $, which for the narrow wormhole corresponds to a small positive Misner-Sharp mass $ M = r _ {\min} / 2 $. Thus, the Misner-Sharp mass first decreases to a large negative value, then returns to the positive region, due to the contribution of the exotic fluid contained in the model. In the symmetric solution considered in \cite {wrmh-rdm}, the $A$-function has a deep minimum in the throat and in the redshift characteristics the wormhole RDM solution behaves in the same way as truncated in the point~3 solution of RDM model without wormhole.

As aside note, the use of RDM model for the description of the dark matter halos is justified even in the case when the matter is relativistic, NRDM. In \cite {static-rdm}, all three types of matter were considered, massive, null and tachyonic, and for them identical asymptotically flat rotation curves were obtained. These curves are characterized by a single parameter $\epsilon=(v/c)^2$, which for all types of matter can be set to a small value, to reproduce the observed non-relativistic velocities of stars in the galaxy. The interior solutions for different types of matter also appear to be similar. The reason for this independence on the type of matter has been explained in \cite {static-rdm}: in the outer region the matter terms contribute only to the slowly varying common factors, while in the inner regions their contribution vanishes. 

Amazingly, Barranco et al. \cite {13016785} obtain different rotation curves for the hot dark matter than for the cold one. Indeed, considering for simplicity in (10) and (14) in \cite {13016785} a limit of flat rotation curves $a\to0$, one obtains a known result with the barotropic equation of state $p=\rho\ (v/c)^2/2$. The measured non-relativistic rotation curves $v\ll c$ correspond to the cold dark matter $p\ll\rho$, rather then the hot one. 

The detailed analysis resolves this controversy: \cite {13016785} considers dark matter with three equally distributed pressure components, while in \cite {static-rdm} and in this section the matter has only radial pressure component. It turns out that this difference significantly affects the equations and the resulting orbital velocities. In the next section, we will consider the case of equally distributed pressure and see that the solution of this model actually differs from RDM.

On the other hand, the purpose of this work is not to study the dark matter models, but to find stable variants of white holes. Here we actually built such a variant, white hole alike massive compact object, permanently radiating outgoing null shells, as well as absorbing ingoing null shells, and remaining stable during arbitrary periods of time. Crossing null shells give rise to the phenomenon of mass inflation, which leads to the formation of extremely massive coat surrounding the naked singularity. As in the previous section, the solution requires the negative mass in the center.

\section{Interaction of white hole with photon gas}\label{sec4}

\begin{figure}
\begin{center}
\includegraphics[width=0.45\textwidth]{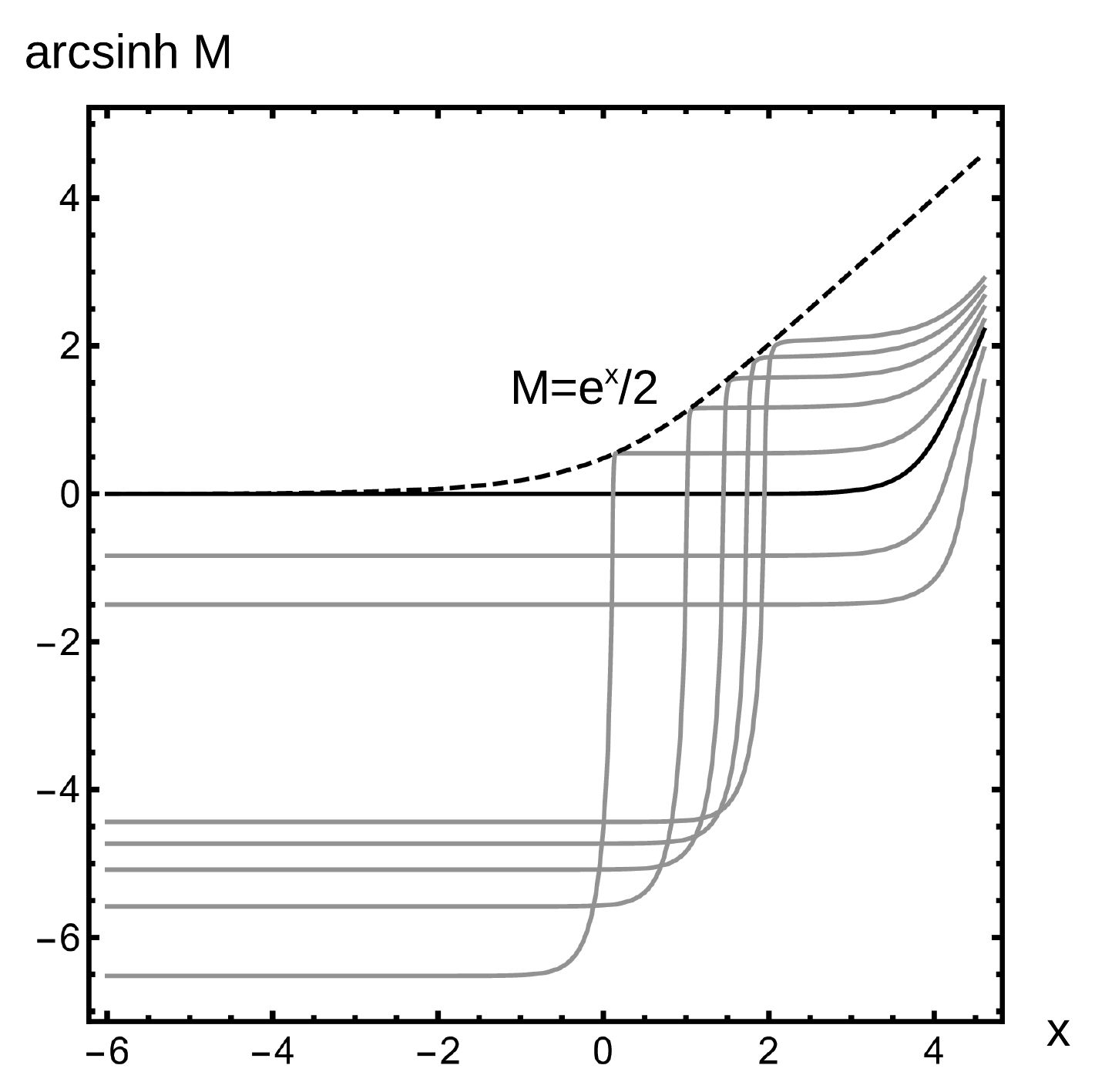}
~~\includegraphics[width=0.45\textwidth]{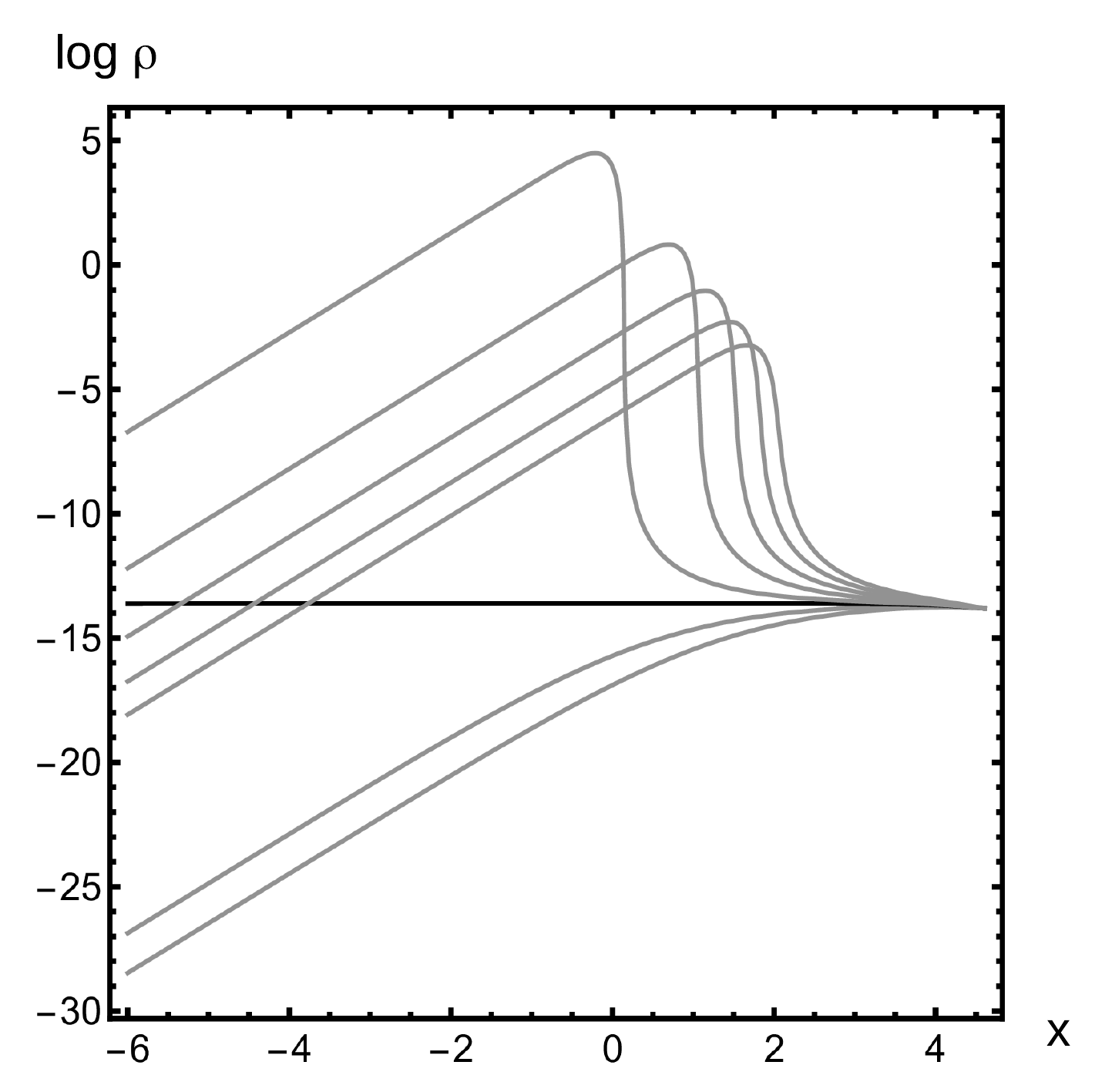}
\end{center}
\caption{On the left: solutions of TOV equations for $w=1/3$, $r_1=100$, $\rho(r_1)=10^{-6}$ and different values of $M(r_1)$. A regular solution, shown by black solid line, corresponds to $M(r_1)=4.5354$ and $M(0)=0$. Solutions below this line correspond to negative $M(0)$ and naked central singularity. Solutions above this line are would-be black holes, they initially tend to positive $M(0)$, but really bounce off the horizon line $M=r/2$, go through mass inflation and end in the central singularity with even larger negative $M(0)$. On the right: the corresponding density curves. The regular solution corresponds to the almost constant density, below this line the solution is rarefied by the action of the central singularity, above this line the density strongly increases when solution approaches the horizon, then is rarefied by the central singularity.}\label{f3}
\end{figure}

In this section, a photon gas in a spherical container with a mirror wall (in a thermos) will be considered. In the center there will be a massive compact object interacting with this gas. Ideally, on the outer border, this system should be stitched with a dynamic cosmological model. Here, as a simplification, we consider the stationary problem, and the mirror wall on the outer boundary will support this stationarity.

The energy-momentum tensor has the form $T^\mu_\nu=\diag(-\rho,p,p,p)$, with isotropically distributed pressure components and the equation of state (EOS) of the form $p=w\rho$, for photon gas $w=1/3$.

The system to solve is Tolman-Oppenheimer-Volkoff (TOV) equations, see, e.g., Blau \cite{Blau} (23.86)+(23.87)+(23.80), in the system of units $G=c=1$:
\begin{eqnarray}
&&w \rho'_r=-(\rho M/r^2)(1+w)(1+4\pi r^3 w\rho/M)(1-2M/r)^{-1},\label{tov1}\\
&&M'_r=4\pi r^2\rho,\ h'_r=4\pi r(1-2M/r)^{-1}\rho(1+w),\label{tov2}
\end{eqnarray}
where $M$ is the Misner-Sharp mass and $\rho$ is the mass density. The system defines the spherically symmetric stationary metric of the form (\ref {stdmetr}) with coefficients
\begin{eqnarray}
&&A=e^{2h}f,\ B=f^{-1},\ f=1-2M/r.
\end{eqnarray}
We solve this system numerically for different starting values $M_1$, $\rho_1$ at the outer radius $r_1$. The solution is presented in the logarithmic coordinates $a=\log A$, $b=\log B$, $x=\log r$. To represent the mass and density behavior at different scales, $\log\rho$ is used as well as $\arcsinh M$, possessing the logarithmic asymptotics at negative and positive infinities.

Fig.\ref{f3} left presents solutions of TOV equations for $w=1/3$, $r_1=100$, $\rho(r_1)=10^{-6}$ and different values of $M(r_1)$. There is a regular solution, shown by black solid line, starting at $M(r_1)=4.5354$ and ending at $M(0)=0$. The solutions starting at smaller $M(r_1)$, have negative $M(0)$, corresponding to naked central singularity. The solutions starting at larger $M(r_1)$, initially behave like they would have positive $M(0)$, however, they inevitably approach the horizon line $M=r/2$, from which they bounce off towards negative masses and end in the central singularity with even larger negative $M(0)$. Thus, the regular solution line separates solutions with naked central singularity from those having this singularity covered by a massive coat, similar to the earlier considered RDM solutions. We deliberately keep off the region $M_1>r_1/2$, where the starting point is located inside the black or white hole.

The density curves, shown on Fig.\ref{f3} right, show that the regular solution in logarithmic scale has almost constant density and the mass function close to $M\sim r^3$. The same plot in the normal scale shows a gradual 22\% increase of density to the center and corresponding slight deviation of the mass function, as needed for the hydrostatic equilibrium. Below the regular solution line the density is rarefied by the action of the central singularity, above this line the density strongly increases when solution approaches the horizon, then is rarefied by the central singularity.

\begin{figure}
\begin{center}
\includegraphics[width=0.45\textwidth]{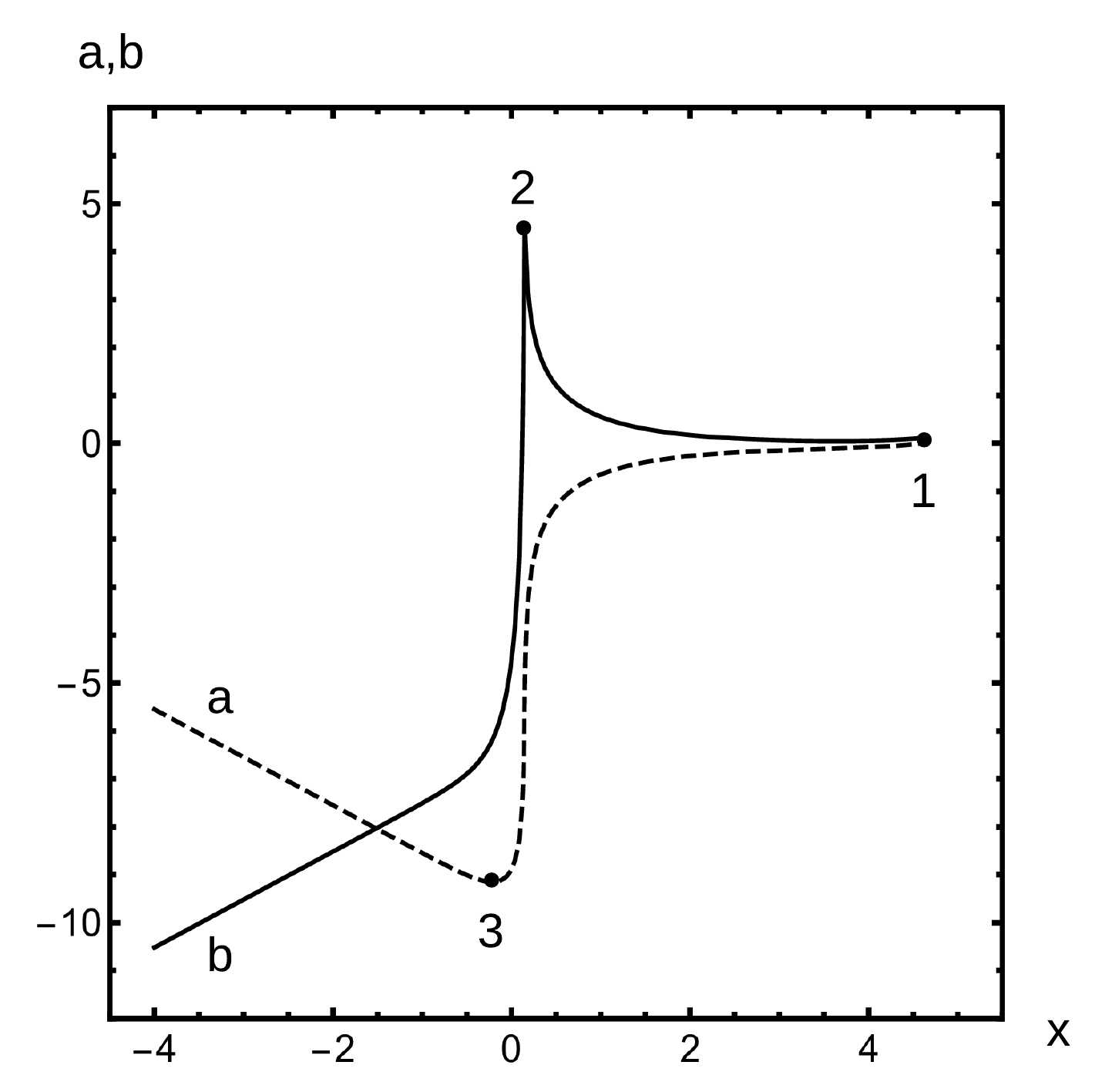}
~~\includegraphics[width=0.45\textwidth]{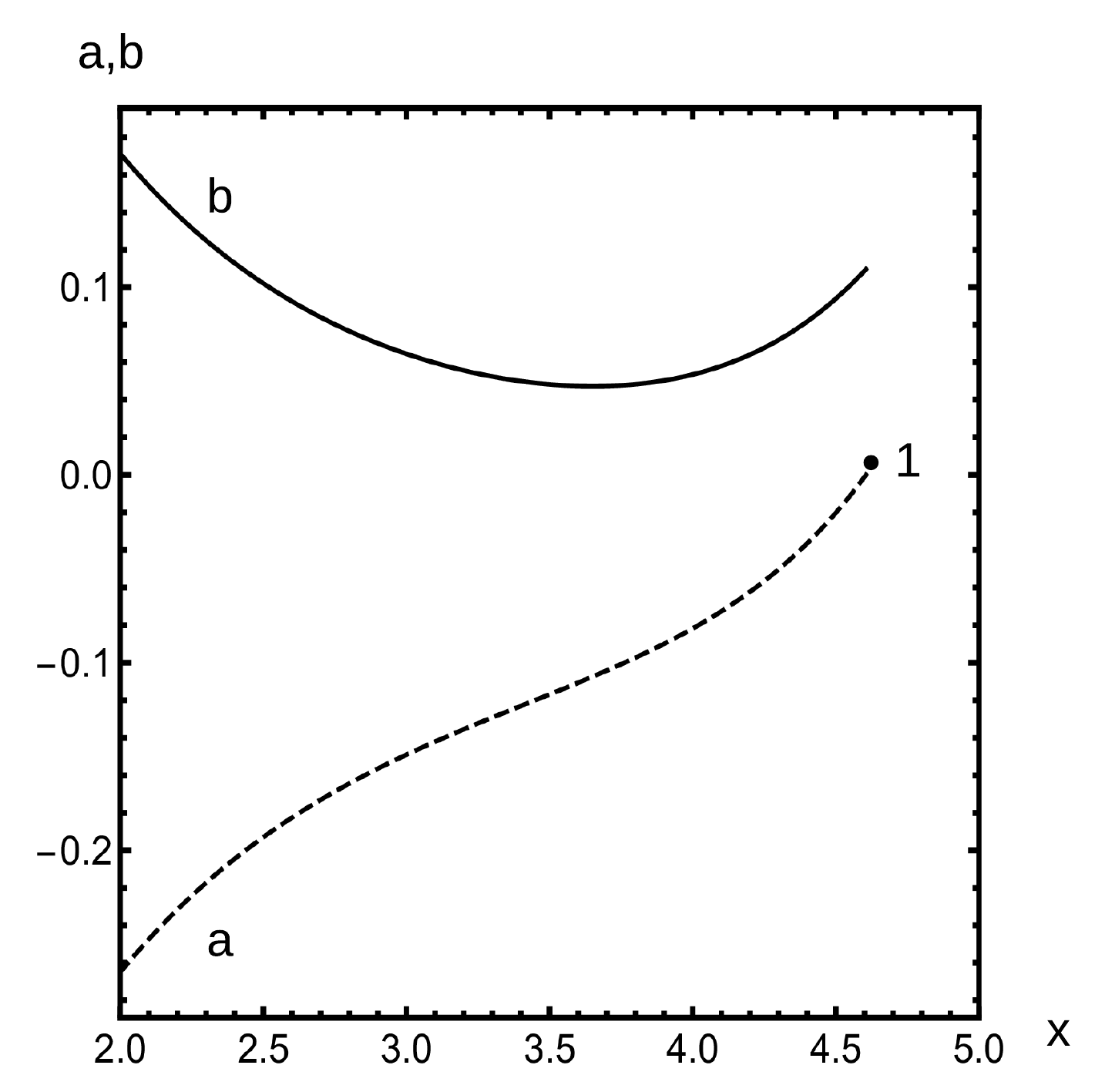}
\end{center}
\caption{Solutions of TOV equations for the same parameters as in the previous figure and $M(r_1)=5.1888$. On the left: $a,b$-functions, on the right: a closeup. The curves show similar behavior as RDM solutions, except of a different asymptotics near the point~1. }\label{f4}
\end{figure}

Fig.\ref{f4} shows $a,b$- and $M$-functions for a selected solution with the starting value $M(r_1)=5.1888$. It can be found in the previous Fig.\ref{f3} as the first gray line over the solid black regular solution. The $a,b$-profiles are similar to RDM solutions. The exception is shown in a closeup on the right, a different asymptotics near the point~1. 

The considered solutions possess large distance asymptotics
\begin{eqnarray}
&&a=Const+4\pi(1+3w)\rho_1 r^2/3 - 2M_0/r,\label{tovasy1}\\
&&b=8\pi\rho_1 r^2/3+2M_0/r,\label{tovasy2}
\end{eqnarray}
different from (\ref{eq_z1}) for RDM model. 
It corresponds to the asymptotically constant density $\rho=\rho_1$, the Misner-Sharp mass $M=4\pi\rho_1 r^3/3+M_0$ and the metric coefficient $b=-\log(1-2M/r)$ at $2M/r\ll1$. The coefficient $a$ in the weak field limit determines the gravitational potential $\varphi=a/2$, which corresponds to the radial gravitational acceleration of the form $v^2/r=\varphi'_r=M_{\mbox{\footnotesize grav}}/r^2$, where $v$ is the orbital velocity, $M_{\mbox{\footnotesize grav}}=4\pi(1+3w)\rho_1 r^3/3+M_0$ is the effective gravitating mass. Pay attention to the multiplier $(1+3w)$, distinguishing this mass from the Misner-Sharp mass. It arises from the fact that not only energy of radiation but also three pressure components contribute to gravity. We also note that with the prevalence of the first term in the gravitating mass over the second, the rotation curves turn out not flat $v=Const$, but linearly growing $v\sim r$.

\begin{table}
\begin{center}
\caption{TOV model scenario with a stellar mass compact object in cosmic microwave background}\label{tab3}

~

\def\arraystretch{1.1}
\begin{tabular}{|c|c|}
\hline
model parameters&$M_1=10 M_\odot$, $w=1/3$, $\rho_1=\rho_{cmb}=4\cdot10^{-14}$~J/m$^3$\\ \hline
starting point of&$r_1=10^6$m, $a_1=0$, $b_1=0.0299773$,\\ 
the integration& $M_1/M_\odot=10$ \\ \hline
&$r_{2}=29532.4$m, $a_{2}=-54.2719$, $b_{2}=53.7265$, \\ 
supershift begins&$M_2/M_1-1=-3.64729\cdot10^{-23}$,\\
& $r_2-2GM_2/c^2=1.37139\cdot10^{-19}$m\\ \hline
&$r_{3}=20638.1$m, \\
supershift ends&$a_{3}=-107.522$, $b_{3}=-104.685$,\\ 
& $\log_{10}(-M_3/M_\odot)=46.3087$\\ \hline
minimal radius& $r_4=1.62\cdot10^{-35}$m, \\ 
(Planck length),&$a_4=-17.6594$, $b_4=-195.278$,\\
end of the integration& $M_4=1.728\ M_3$\\ \hline
\end{tabular}

\end{center}
\end{table}

For numerical integration of realistically dimensioned scenarios, it is more convenient to present the system in the form, that was used in our previous work \cite{wrmh-rdm} for the description of an exotic fluid in the RDM model of wormholes. 

First of all, the consequence of the TOV system is the hydrostatic equation
\begin{eqnarray}
&&r (p+ \rho) A'_r + 2 A r p'_r =0
\end{eqnarray}
for our EOS possessing general solution
\begin{eqnarray}
&&4\pi w\rho=k_3 A^{k_4}.
\end{eqnarray}
In comparison with \cite{wrmh-rdm}, factor $4\pi$ comes from different system of units. We also write all constants defined in \cite{wrmh-rdm} here for compatibility:
\begin{eqnarray}
&&k_1=1/w,\ k_2=1,\ k_3=4\pi\rho_1 w,\\
&&k_4=-(1 + k_1)/2,\ k_5=0,\ k_6=\log k_3.
\end{eqnarray}
We use constant $k_3>0$, differently from \cite{wrmh-rdm} where the exotic fluid with $k_3<0$ was considered.

Then, TOV system becomes equivalent to 
\begin{eqnarray}
&&a'_x=-1 + e^b + 2 e^{2 x + k_4 a + b + k_6},\label{tovax}\\ 
&&b'_x= 1 - e^b + (2/w)e^{2 x + k_4 a + b + k_6}\label{tovbx}
\end{eqnarray}
with initial data $a_1=0$, $b_1=-\log(1-2M_1/r_1)$, where $w=1/3$, $k_4=-2$, $k_6=\log(4\pi\rho_1/3)$, $\rho_1$ and $M_1$ at $r_1$ are given.

Before numerical integration, it is better to normalize the system
\begin{eqnarray}
&&a'_y=v_a/norm,\ b'_y=v_b/norm,\ x'_y=v_x/norm,\\
&&norm=\sqrt{v_a^2+v_b^2+v_x^2},
\end{eqnarray}
where $v_a$, $v_b$ are the right hand sides of (\ref{tovax}),(\ref{tovbx}) and $v_x=1$. The solution is then given parametrically as $(a,b,x)(y)$. The integration is performed with {\it Mathematica} algorithm {\tt NDSolve}. The critical points of solution are presented in Table~\ref{tab3}.

Here we consider a compact object of stellar mass $M_1=10M_\odot$, with a nominal gravitational radius of about $r_s\sim30$~km, placed in cosmic microwave background of energetic density $\rho_{cmb}=4\cdot10^{-14}$~J/m$^3$. Integration begins at a distance $r_1=1000$~km. At larger distances, the density of the photon gas is almost constant, and nothing interesting happens.

The maximum value of the metric coefficient $b_2\sim54$ is reached at a distance close to the gravitational radius $r_2\sim r_s$. Note the extremely small gap $r_2-r_s\sim10^{-19}$~m separating the object from the gravitational collapse. This is the distance at which initially inactive matter terms, describing relic radiation, wake up and begin to influence strongly the structure of the solution. They are responsible both for the finite value at the maximum of $b$-coefficient, and for the subsequent phenomenon of mass inflation, which proceeds similarly, but not so stiff as in the RDM model. 

The minimum value of the coefficient $a\sim-108$ is reached at the point $r_3\sim21$~km. The mass $M_3/M_\odot=-2\cdot10^{46}$ reached there is extremely large negative, but still significantly less than the mass values reached in the RDM model. Further, up to the Planck length, the mass slightly grows in the negative direction $M_4/M_\odot=-3.5\cdot10^{46}$, while the $a,b$-factors behave as they should be for the naked Schwarzschild singularity, $a$ grows, $b$ decreases. The peculiarity here is the presence of a strong dip (supershift) in the initial values of the coefficients, as a result of which the coefficient $a$ grows till the Planck length only to the value of $a_4\sim-18$ located in the infrared region. This value corresponds to the moderate redshift factor $A_4^{-1/2}=\exp(-a_4/2)\sim6.8\cdot10^3$. 

Compared to the naked Schwarzschild singularity $A_4=1-2M_4/r_4\sim\exp(-b_4)$, for the obtained extremely large negative mass the extremely high ultraviolet shift factor would be reached $A_4^{1/2}=\exp(-b_4/2)\sim2.5\cdot10^{42}$. To illustrate the strength of this ultraviolet shift, consider one photon with typical energy of the background radiation $T=2.7$~K, with Boltzmann factor $kT\sim3.7\cdot10^{-23}$~J, after ultraviolet shift it will be $\sim10^{20}$~J, which corresponds to a mass of about 1~ton. 

A distant observer bombarded by such photons has a good reason to fear for safety. This is why naked singularities of negative mass are considered as dangerous cosmic objects. To our common happiness, in reality, ultraviolet accelerators of such power do not appear, in full accordance with the principle of cosmic censorship. On the other hand, in the scenario considered here, the super-strong ultraviolet shift is compensated by super-strong infrared one. The remote observer receives the relic photon {\it weakened} in energy by $\sim$7~thousand times, giving the cosmic censor no reason to impose a ban.

For comparison, we also performed the calculation for TOV model with the Milky Way parameters: $r_s=1.2\cdot10^{10}$~m, $M_1/M_\odot=4.06\cdot10^6$, $w=1/3$, $\rho_1=\rho_{cmb}$, $r_1=100r_s$. The result is a solution of a similar structure, on the Planck length having ultraviolet shift with a moderate factor $\exp(a_4/2)\sim3.8\cdot10^4$. Compared to the naked Schwarzschild singularity, this ultraviolet factor effectively raises the radiation temperature just to $T\sim100$~thousand K, rather cool by cosmic standards.

At large distances, the mass density in the considered scenarios is approximately constant. Integration of the stellar mass scenario outwards of $r_1=10^3$~km shows that $\rho$ quickly goes to a constant and between $r\sim5.4\cdot10^3$~km and $r\sim2.6\cdot10^7$~km changes only by $\Delta\rho/\rho\sim1$\%.

\paragraph{Remark on self-similar solution and isothermal halo.} The structure of solutions described above represents only a part of the big picture. TOV equations (\ref{tov1}), (\ref{tov2}) are invariant with respect to the scaling transform
\begin{eqnarray}
&&r\to cr,\ \rho\to c^{-2}\rho,\ M\to cM.\label{scaltov}
\end{eqnarray}
This means that this transform translates the solutions of the system into its solutions, generally others. At the same time, the work of Visser and Yunes \cite{0211001} showed that TOV system possesses a fixed point for associated autonomous equation, and, in fact, TOV system possesses a solution that, being considered as a curve in $(r,\rho,M)$ space, is translated by scaling transform (\ref {scaltov}) into itself. This scale-invariant or self-similar solution has the form
\begin{eqnarray}
&&\rho=(2w/(1 + 6 w + w^2))/(4\pi r^2),\\
&&M=(2w/(1 + 6 w + w^2))r,\\
&&h=Const+(2w/(1+w))\log r,
\end{eqnarray}
corresponding to metric coefficients
\begin{eqnarray}
&&a=Const+(4 w / (1 + w))x,\\ 
&&b=\log((1 + 6 w + w^2)/(1 + w)^2),
\end{eqnarray}
$a$ is a linear function of $x$, $b$ is a constant.

Near the critical point, where scale-invariant solution appears, complex phenomena, bifurcations occur. In particular, at low density $ \rho_1 $, the boundary problem $ M(0) = 0 $ has only one solution, i.e., only one initial $ M_1 $ leads to the final $ M(0) = 0 $. As our numerical experiments show, near the critical point this problem has several solutions, that is, at a fixed boundary density, there are several regular solutions separating naked singularities from solutions with the massive coat. The complex structure of these alternating singular and regular solutions deserves special consideration. On the other hand, we will see below that the scenario of interest to us is located far from scale-invariant solution, in a subcritical regime.

The scale-invariant solution of TOV system physically corresponds to the so-called isothermal halo in the model of dark matter in spiral galaxies. Using the exact relativistic formula for orbital velocity from \cite{static-rdm}
\begin{eqnarray}
&&v^2=a'_x/2=2 w/(1 + w), \label{relorb}
\end{eqnarray}
we see that for any $w$ we obtain constant (flat) rotation curves. In Barranco et al. work \cite {13016785} for plane rotation curves and small $w$, a non-relativistic formula $v^2\sim2w$ consistent with (\ref{relorb}) was obtained. For $w=1/3$, the relativistic orbital velocity $v^2=1/2$ is obtained. From this we can conclude, in agreement with \cite {13016785}, that the experimentally observed non-relativistic orbital velocities are reached only for small $w$, cold dark matter, when describing it by the scale-invariant TOV solution. In this aspect TOV solution differs from RDM, in which all types of matter produce asymptotically flat rotation curves with freely adjustable factor $ \epsilon $. In particular, the NRDM model can be configured to obtain non-relativistic flat rotation curves.

Considering this question in even more detail, in the limit of more and more rarefied gas, $\rho_1 \to0$, TOV will produce the asymptotics (\ref{tovasy1}), (\ref{tovasy2}), with constant density and linear rotation curves $v\sim r$. Note that RDM will always produce $\rho\sim r^{-2}$, due to the geometry of the system with radially convergent flows of matter. This is exactly the density profile that is required for flat rotation curves.

On the other hand, if we fix TOV and require, for agreement with the experiment, asymptotically flat rotation curves, then their realization can be achieved on the scale-invariant solution, whose formation requires the critical density. This density is proportional to $w$, thus small densities and non-relativistic orbital velocities at the scale-invariant TOV solution are available only for small $w$. In RDM case, the rotation curves are always asymptotically flat, and the orbital velocities can be freely adjusted from non-relativistic to relativistic by simple scaling of density. The reason for such different behavior is the inclusion of EOS with tangential pressure components, this changes the type of equations and the structure of their solutions.

There is another, methodological difference between our work and \cite{13016785}. Although the same TOV system was solved, in \cite{13016785} this system was not solved with respect to the metric coefficients for a given EOS. The problem was solved as if from the other end, substituting the known rotation curves into the equations and finding EOS from them. In this approach, differential equations are solved by direct integration of experimentally known profiles. At the same time, the question that interests us, what happens deep inside the system, remains unanswered, since there are no experimental profiles there.

For clarity, let us consider several scenarios in which we will evaluate the critical mass and density for a given external radius:
\begin{eqnarray}
&&\rho_{crit}=(2w/(1 + 6 w + w^2))/(4\pi r_1^2),\\
&&M_{crit}=(2w/(1 + 6 w + w^2))r_1.
\end{eqnarray}

At $w=1/3$, $r_1=1$~m, $M_{crit}=0.21$~m, in physical units it corresponds to $3\cdot10^{26}$~kg, 15\% Jupiter mass converted to radiation and closed in the container of 1~m radius. 

Further, consider non-relativistic ideal gas, $w=RT/(\mu c^2)$. For definiteness, fix parameters of nitrogen $N_2$, $\mu=28\cdot10^{-3}$~kg/mol, $T=273.15$~K, $w=9\cdot10^{-13}$. For $r_1=1$~m obtain $\rho_{crit}=2\cdot10^{14}$~kg/m$^3$, which is 14 orders of magnitude greater than the density of nitrogen at normal pressure and temperature. So one should not worry that the gas in the balloon will start to form black holes or scale-invariant solutions.

Considering again $w=1/3$, at $r_1=3.1\cdot10^{21}$~m, the radiation in the volume of Milky Way. The critical case corresponds to the energetic density $\rho_{crit}c^2=0.21$~J/m$^3$, which is 13 orders of magnitude greater than $\rho_{cmb}=4\cdot10^{-14}$~J/m$^3$. Thus, the considered problem is in a deeply subcritical regime.

In this regime, at large distances, $\rho$ is almost constant, TOV equations can be used to find the next order correction
\begin{eqnarray}
&&\rho = \rho_1 + \rho_1^2(r_1^2-r^2)\, 2\pi (1 + w)(1 + 3 w)/(3w) +...
\end{eqnarray}
This function defines the density bump required in the first non-vanishing order for hydrostatic equilibrium. In this formula, the second term is much smaller than the first, when the regime is subcritical.

\paragraph{Remark on naked singularities and cosmic censorship.} In the studied models, the central singularities are not covered by event horizons and, formally speaking, are naked. A photon from the singularity in principle can reach the distant observer. However, there is something instead of event horizon -- the supermassive coat, that provides an extremely strong redshift for this photon. The coat is thicker for RDM model and thinner for TOV. In RDM model, if the photon, escaping from Planck’s vicinity of the singularity to infinity, has the initial energy reasonably bounded from above, the final energy will be extremely small, making this photon practically unobservable. In TOV model, the final shift is moderate, balanced between infrared and ultraviolet, dependently on the model parameters. In any case, the final energy is strongly suppressed comparing with the case, when the singularity of the same mass would be really naked. The coat weakens the extremely strong ultraviolet effect arising from gravitational repulsion from very small distances for very large negative masses, which is typical for the time-like singularity considered in this paper.

There are also singularities of spacelike type. This behavior has a white hole in Schwarzschild model, in which the central singularity is naked, since it is covered by particle horizon, not by event horizon, and the light from it reaches the external observer. Zeldovich et al. \cite{Zeldovich} considered a dynamic problem with a white hole having the positive central mass and the spacelike singularity, in contrast to the stationary models we considered, with the negative central mass and the timelike singularity. In scenario \cite{Zeldovich}, the white hole is individualized from the surrounding space in a finite relatively short time, during which the white hole can explode and throw out the matter. After this time, the white hole does not explode, and the matter remains under the horizon. The white hole actually becomes black, encapsulated from the rest of the universe. Apparently, this model describes the processes for which the starting point on the diagram Fig.\ref{f3} is located above the horizon line, the solution is bounced off the horizon from the inside and subsequently remains under the horizon. This is the difference with the timelike singularity, for which the solution is located outside the horizon and extends to infinity.

\begin{table}
\begin{center}
\caption{Considered models and their symmetries}\label{tab4}

~

{\footnotesize 
\begin{tabular}{|c|c|c|c|}\hline
model/scenario & T-symmetry & stationarity & isotropy and homogeneity\\ 
& & & at large distance\\ \hline
null shell, generic & $-$ & $-$ &$-$ \\ \hline 
null shell, T-symmetric & $+$ & $-$ &$-$ \\ \hline 
RDM & $+$ & $+$ &$-$ \\ \hline 
TOV $M_1=M_{crit}$, $\rho_1=\rho_{crit}$ & $+$ & $+$ &$-$ \\ \hline 
TOV $M_1\ll M_{crit}$, $\rho_1\ll\rho_{crit}$ & $+$ & $+$ &$+$ \\ \hline 
\end{tabular}
}

\end{center}
\end{table}

\section{Conclusion}

In this paper, we studied the possibility of stabilizing white hole models by more realistic modeling of external matter falling on the white holes, as well as introducing a core of negative mass into the model.

First of all, we investigated the question, what will happen if one directs a converging null shell from outside to a white hole emitting a diverging null shell. In the standard scenario, if one throws a null shell of initially low energy into a white hole, then waits 13.8~billion years during which the shell hangs on the particle horizon and strengthens itself by ultraviolet shift, then it collides with the outgoing shell, acting as an opaque wall, letting practically nothing out. As we have shown, the model also has other solutions. In particular, the white hole can emit an amount of energy greater than its initial mass, so large that the outgoing shell can break through the almost opaque wall created by the incoming shell. The core of the negative mass remains behind, which is then compensated by the incoming shell and exists is only a finite time in the transition process, so that at both time infinities there are only positive masses.

In an alternate scenario, a white hole radiates the null shells continuously and also absorbs the incoming shells continuously falling on it. This solution is T-symmetric and stationary. The intersection of flows leads to mass inflation phenomenon, as a result of which the removal of the particle horizon and the event horizon occurs, and a massive compact object is formed, almost reaching its gravitational radius, but not crossing it. Inside the object there is a massive coat surrounding the singularity of negative mass. The strong ultraviolet shift in the Planck neighborhood of the singularity is compensated by the strong infrared shift from the coat, as a result, the photons born near the singularity, reasonably limited in the initial energy, reach the distant observer with an extremely small final energy.

We also investigated the question, what will happen if we replace the converging and diverging shells in this scenario with a real photon gas. In gas, photons move in all possible directions, in radially converging, in radially diverging, as well as in tangential ones. In this case, in the subcritical regime, at large distances, the complete symmetry of the system is restored, including isotropy, uniformity, stationarity and T-symmetry. Table~\ref{tab4} describes the models considered in this paper, where the subcritical TOV gas has maximum symmetry. Inside, such a solution looks similar to a solution with radial flows of matter, it also has a massive inflation coat surrounding the central singularity of negative mass. Just the dependencies in this solution turn out to be less sharp. In particular, the shift of photons from Planck neighborhood is moderate and balances between infrared and ultraviolet, depending on the choice of model parameters.

Formally speaking, the stationary solutions considered here are not white or black holes in the exact sense, since they do not have event horizons or particle horizons. They are similar to quasi-black holes, gravastars, fuzzballs, bosonic stars, other dark stars, reviewed by Visser et al. in \cite{09020346}, because outside of the gravitational radius the solution can be mathematically as close as one likes to Schwarzschild black hole, although inside the solution is arranged quite differently. The objects considered in our work exhibit the properties of white and black holes at the same time, they erupt matter and absorb matter, remaining stable for an unlimited time. Note that the matter ejected and absorbed by these objects can be dark, as in the halo model for spiral galaxies, then these objects will look like dark stars almost indistinguishable from black holes. This matter can also be formed by ordinary photons or other relativistic particles from the normal matter sector. Since the most realistic model is the subcritical TOV solution, the deviation of photon gas density caused by these objects becomes large only in the immediate vicinity of the object, while at long distances the photon gas is homogeneous and isotropic without detecting the presence of compact massive objects in it. These objects can also be identified by the gravitational lensing of the light rays and the orbital velocity of the celestial bodies captured by them, and due to the similarity of the external metric to the Schwarzschild one, these objects will be indistinguishable from the ordinary black holes. Thus, in a stationary, equilibrium state, these objects successfully mimic the black holes, like the other objects described in \cite{09020346}.

A feature of the objects we studied is the presence of a central timelike singularity of negative mass and the existence of a light trajectory connecting it with a remote observer. Thus, if in the depths of the object, near the singularity of negative mass, any dynamical processes will occur, signals about them can reach a remote observer, shifted in frequency to infrared or ultraviolet, in the form of radio or gamma bursts. The specific signature of these bursts depends on the exact model of the process, and additional investigation will be required to clarify it.

Another characteristic feature of the studied objects are the huge, almost compensating each other, masses of the exotic core and the inflation coat, in absolute value significantly exceeding the mass of the observable universe. This may mean that the studied structure corresponds to a theoretical stationary limit, which is practically not reached or takes a lot of time to reach. A hypothetical mechanism for the appearance of an exotic core can be the dissociation of matter into particles of positive and negative mass that occurs at superhigh energies. If such a process takes place long enough, it can lead to the formation of an equilibrium configuration of the exotic core and inflation coat. At normal energies, this mechanism can be suppressed, for example, by a mass threshold, if the process goes through the formation of intermediate supermassive particles, and is activated only at high energies. It would be interesting to perform the calculations for the corresponding dynamic scenario.

\end{document}